# A Survey of the Design Trade Space for Atmospheric Entry, Descent, and Landing Missions

Athul Pradeepkumar Girija [1***]

[1]*School of Aeronautics and Astronautics, Purdue University, West Lafayette, IN 47907, USA*

## ABSTRACT

Over the history of planetary exploration, atmospheric entry vehicles have been used to deliver probes and landers to Venus, Mars, Jupiter, and Titan. While the entry vehicles are tools for furthering scientific exploration, by delivering probes and landers which perform in-situ exploration, the entry vehicle and trajectory design in itself is of significant interest. Entering an atmosphere subjects the vehicle to deceleration and aerodynamic heating loads which the vehicle must withstand to deliver the probe or the lander inside the atmosphere. The conditions encountered depend on the destination, the atmosphere-relative entry speed, the vehicle type, ballistic coefficient, the vehicle geometry, and the entry-flight path angle. The driving constraints are the peak aerodynamic deceleration, peak heat rate, and the total heat load encountered during the critical phase of the entry. Jupiter presents the most extreme entry conditions, while Titan presents the most benign entry environment. This study presents a survey of the design trade space for atmospheric entry missions across the Solar System using carpet plots, along with benchmarks from historical missions, which can serve as a useful reference for designing future missions.

---

**** To whom correspondence should be addressed, E-mail: athulpg007@gmail.com

# I. INTRODUCTION

Over the history of planetary exploration, atmospheric entry vehicles have been used to deliver probes and landers to Venus, Mars, Jupiter, and Titan. Sample return missions have also used entry vehicles to enter the Earth's atmosphere. The atmospheric probes and landers have been able to make in-situ measurements that cannot be done otherwise, and have been critical to our understanding of the origin and evolution of the Solar System [1]. While the entry vehicles are tools for furthering scientific exploration, by delivering probes, or landers and rovers which perform in-situ exploration, the entry vehicle and trajectory design in itself is of great engineering interest [2]. Entering an atmosphere subjects the vehicle to deceleration and aerodynamic heating loads which the vehicle must withstand to deliver the probe or the lander inside the atmosphere. The conditions encountered strongly depend on the destination's gravity well, atmospheric structure, and composition [3]. Other factors include the atmosphere-relative entry speed; the vehicle type, ballistic coefficient (mass-to-area ratio), the vehicle geometry (cone angle and bluntness, i.e. nose radius), and the entry-flight path angle (steep vs shallow). The objective of the entry vehicle is to deliver the science payload or lander inside the atmosphere at conditions where the instruments can make their measurements or a lander can perform a soft landing. The key design variables available for entry vehicle design are the ballistic coefficient, the vehicle geometry, the entry-flight path angle, and the entry speed [4]. The two common configurations for entry vehicle design are shown in Figure 1. Conventional rigid aeroshells such as the Pioneer Venus and the Galileo entry probes have a high-ballistic coefficient. Deployable aeroshells such as ADEPT have large drag areas, and result in low-ballistic coefficient. The driving constraints are the peak aerodynamic deceleration, peak heat-rate, and the total heat load encountered during the critical phase of the entry. The peak deceleration drives the vehicle structural design, as well as the scientific instruments that can be carried which need to be qualified to withstand the deceleration loads. The peak stagnation point heat rate drives the choice of thermal protection system (TPS) materials which can be used. TPS materials are typically qualified to withstand a combination of heat rate and stagnation pressure in ground based arc-jet facilities. The total heat load drives the TPS mass fraction, and must be minimized to keep the TPS mass fraction low, and allow a reasonable payload mass fraction. Traditionally, entry vehicle design trade space is analyzed using a 'carpet plot', which shows the deceleration, heat-rate, and heat-load as a function of the vehicle ballistic coefficient, entry speed, and entry flight-path angle. This study presents a survey of the design trade space for atmospheric entry missions across the Solar System using carpet plots, along with benchmarks from historical missions, which can serve as a useful reference for designing future entry probe and lander mission concepts.

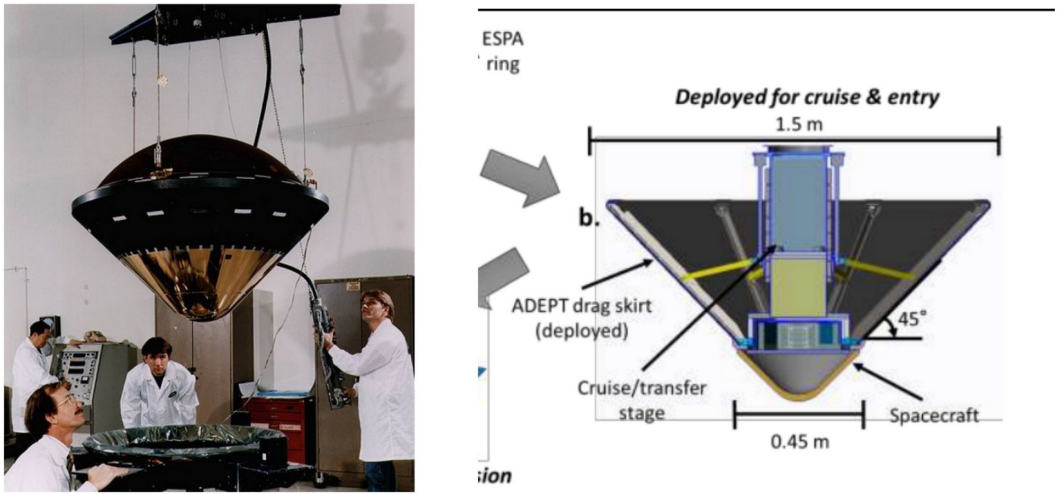

Figure 1. (Left) The Galileo entry probe. (Right) Deployable ADEPT entry system (Austin et al. [9])

The vehicle ballistic coefficient (beta) is one of the most important design parameters for the entry vehicle. All planetary entry vehicles flown to date have been rigid aeroshells, with the exception of some recent suborbital flight tests of deployable and inflatable aerodynamic decelerators at Earth. A high ballistic coefficient subjects the vehicle to higher deceleration and heating rates than a low ballistic coefficient vehicle. Rigid aeroshells such as the Pioneer Venus or Galileo probe have ballistic coefficients in the range of ~100s of $kg/m^2$, whereas deployable low-ballistic coefficient aeroshells such as ADEPT fall in the range of ~10s of $kg/m^2$. The low-ballistic coefficient vehicles are able to decelerate much higher up in the atmosphere compared to rigid aeroshells, where the atmosphere is thinner and thus keep deceleration and heating rates low. The entry speed is largely driven by the planet's gravity well and to a lesser extent by the interplanetary arrival speed. Higher entry speeds result in more demanding entry environments. For example Jupiter's enormous gravity well resulted in the Galileo probe encountering an atmosphere-relative speed of 47 km/s, the highest entry speed ever attempted, and was the most difficult atmospheric entry in the history of space exploration. For fast-rotating outer planets, entering prograde can result in substantially smaller atmosphere-relative entry speeds than retrograde entry. The atmospheric structure and composition play a role in the entry environment encountered. A short, thick atmosphere such as Venus imparts large deceleration and heating loads. An extended, thick atmosphere, along with low entry speeds such as Titan enables conditions that are quite modest compared to Venus. The entry-flight path angle is another important design variable. Entering too shallow can result in the vehicle skipping out, while entering too steep will result in high deceleration and peak heat rate. Entering too shallow can also result in the vehicle encountering heating for an extended duration, causing the heat load (integral of the heat rate over time) to be too high. To keep the TPS mass fraction reasonable, the total heat load must be constrained [5].

## II. SURVEY OF THE DESIGN TRADE SPACE FOR ATMOSPHERIC ENTRY

The Aerocapture Mission Analysis Tool (AMAT) designed for rapid conceptual design of aerocapture and entry missions, is used to generate the carpet plots for the trade space survey [6, 7]. For a selected ballistic coefficient and nose radius, the charts show the peak deceleration, peak heat rate, and total heat load as a function of the entry flight-path angle (EFPA) and the atmosphere-relative entry speed. Figure 2 shows the trade space for Venus entry with a rigid aeroshell such as the Pioneer Venus. The vehicle design parameters are chosen from the Pioneer Venus aeroshell (beta = 190 kg/m$^2$, nose radius RN = 0.19 m). As discussed earlier, both the deceleration and heat rate increase with steeper EFPA and increasing entry speed. The total heat load also increases with entry speed. However, for the same entry speed, the total heat load is higher for shallow entry compared to steep entry. Venus presents a particularly demanding destination for entry due to its short thick atmosphere. The four Pioneer Venus (PV) entry probes are highlighted in Figure 2 for reference. The PV Small North probe for example entered at -68.74 deg, encountering deceleration in excess of 450g and peak heat rate of nearly 7000 W/cm$^2$. The PV Small Day probe which entered much shallower at -25 deg encountered less extreme conditions than the North probe (225g, 3500 W/cm$^2$), but was still quite demanding. Figure 2 shows that Venus entry with rigid aeroshells is quite demanding with decelerations in the 100s of g, heat rates in 1000s of W/cm$^2$, and heat loads in the range of 15 − 30 kJ/cm$^2$.

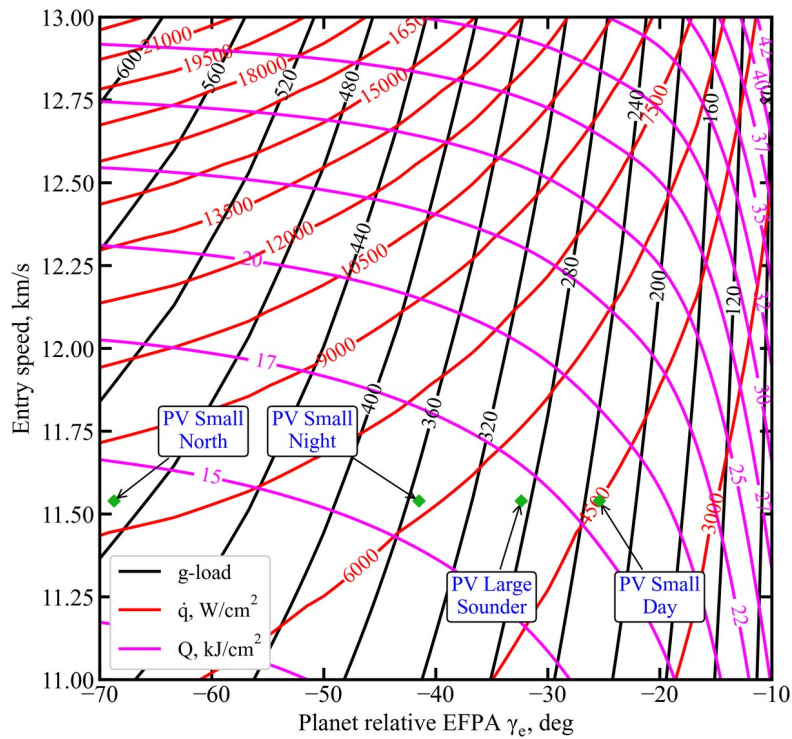

Figure 2. Trade space for Venus entry with rigid aeroshell (beta = 190 kg/m$^2$).

To overcome the problem of high decelerations and heat rates with rigid aeroshells at Venus, low ballistic coefficient vehicles are a promising alternative (Dutta et al., 2012). Compared to rigid aeroshells which have to enter steep to keep the heat load small, the low ballistic coefficient vehicle can enter shallow (just steep enough to avoid skip out) and thus keep the deceleration and heat rates low. Figure 3 shows the trade space for Venus entry with a low ballistic coefficient entry system such as ADEPT. The vehicle design parameters are for a typical 1.5m small ADEPT deployable entry system (beta = 25 kg/m$^2$, nose radius RN = 0.20 m). Note that the EFPA range is much shallower than shown in Figure 1. For an entry speed of 11.5 km/s (same speed as PV entry), and entering at a shallow angle such as -9 deg, the deceleration and peak heat rate can be limited to 60g and 600 W/cm$^2$, and heat load to 10 kJ/cm$^2$. Comparing Figs. 1 and 2, the advantages offered by deployable entry systems compared to rigid aeroshells in terms of lower deceleration and heating rates for Venus missions is evident. The demanding entry environment for high ballistic coefficient systems also make Venus not an attractive target for aerocapture with traditional blunt body aeroshells [8]. Propulsive capture and aerobraking is the recommended method for orbit insertion, and is baselined for all upcoming Venus missions. However, aerocapture using low ballistic coefficient drag modulation systems for inserting small satellites is an attractive option [9]. Low-ballistic coefficient entry system have applications to be used for delivering secondary payloads to Venus [10], small satellites supporting Flagships [11], and sample return missions [12].

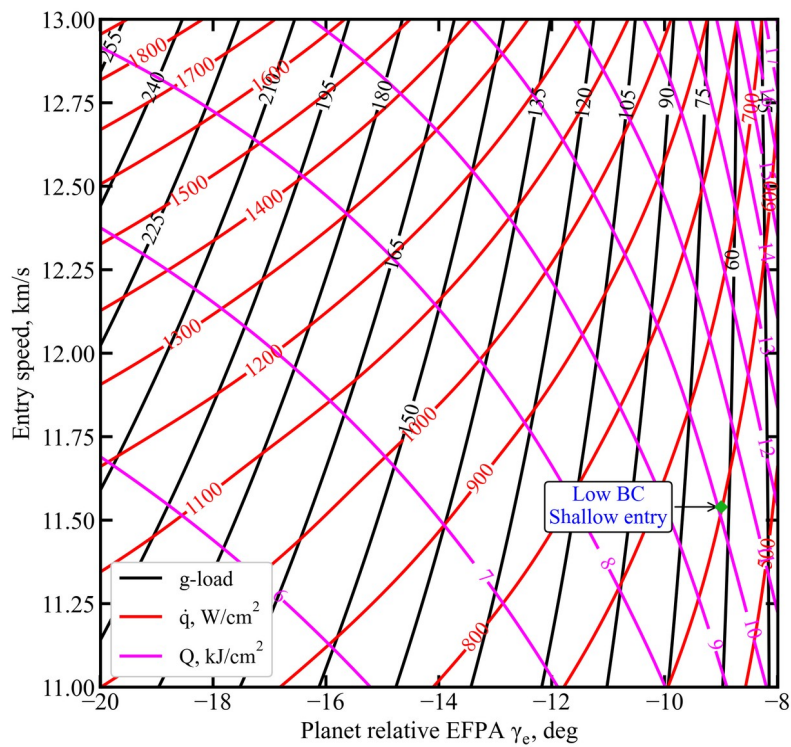

Figure 3. Trade space for Venus entry with low ballistic coefficient system (beta = 25 kg/m$^2$).

Figure 4 shows the trade space for Earth entry for sample return missions with rigid aeroshells (left, beta = 60 kg/m$^2$), and low-ballistic coefficient systems (right). Some historical missions such as Stardust and Genesis, are highlighted for reference. All these missions entered at nearly the same shallow EFPA = -8 deg, and encountered peak heat rates in the range of 600-1400 W/cm$^2$. Figure 5 shows the trade space charts for Mars (beta = 146 kg/m$^2$). In the inner Solar System, Mars offers the least demanding entry conditions with peak heat rates under 100 W/cm$^2$.

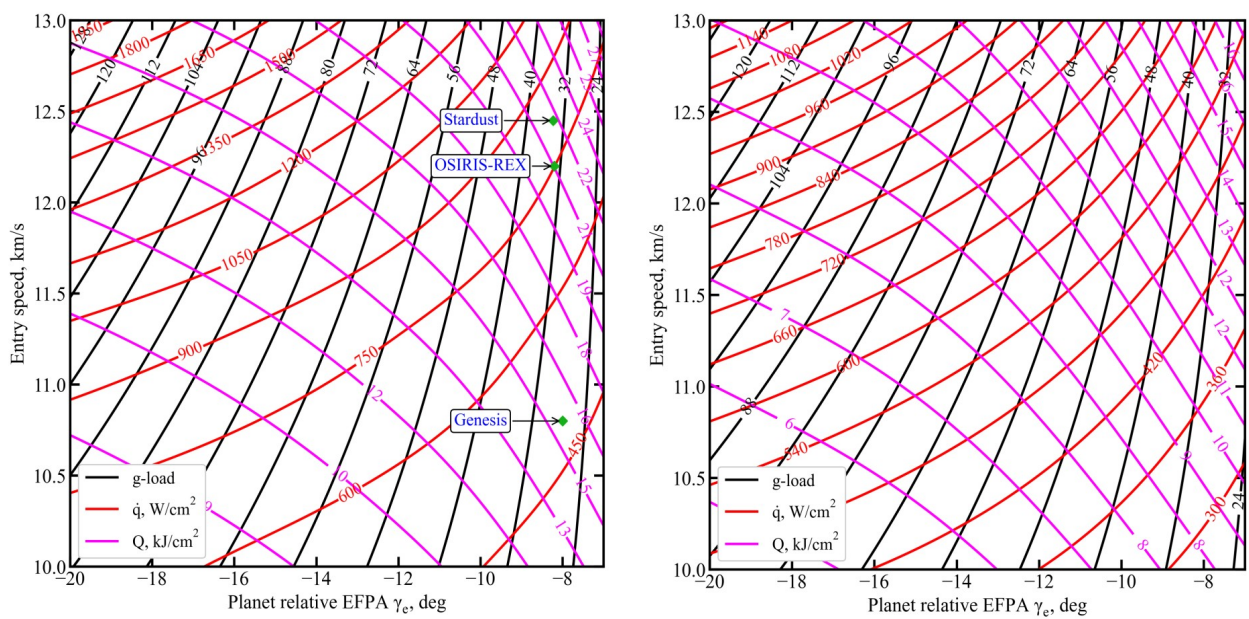

.Figure 4. (Left) Trade space for Earth entry with rigid aeroshell, and (right) low b.c.(beta = 25 kg/m$^2$).

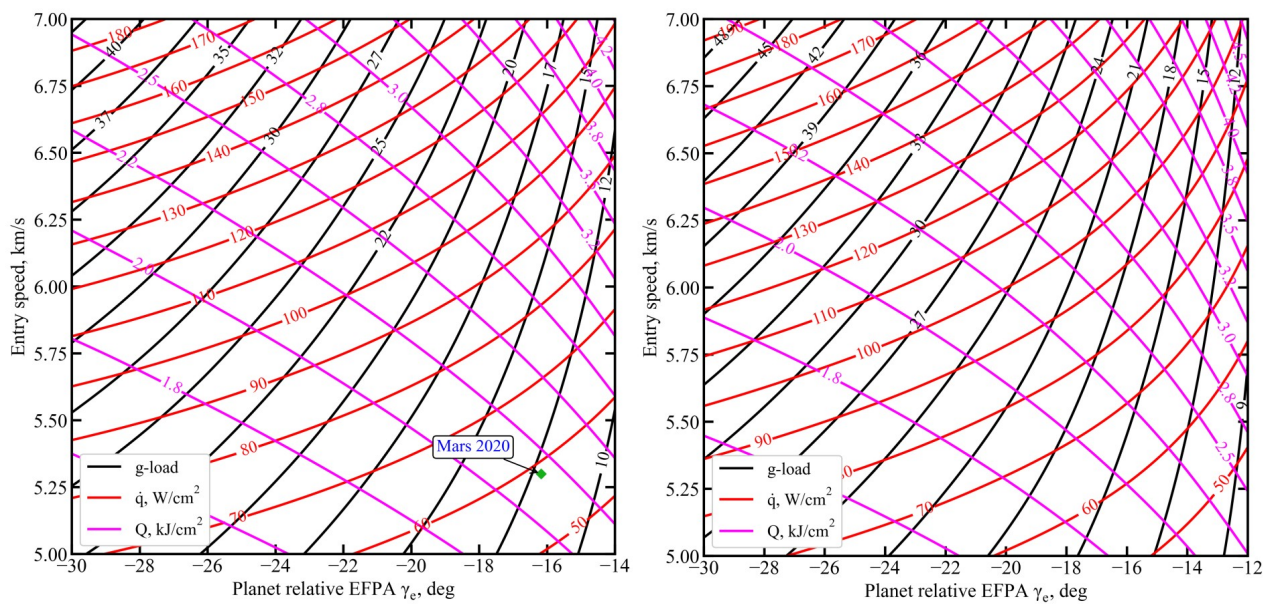

Figure 5. (Left) Trade space for Mars entry with rigid aeroshell, and (right) low b.c.(beta = 25 kg/m$^2$).

Comparing Figs. 1 and 4, it is seen Venus entry using rigid aeroshells is much more demanding than that at Earth, due to the thick Venusian $CO_2$ atmosphere despite the entry speeds being similar. Comparing Figs. 4 and 5, it can be seen that the entry environment at Mars is much less demanding than that at Earth. This is attributed to the thin Martian atmosphere and the lower entry speeds at Mars. The relatively benign entry conditions make Mars a very attractive target for deployable entry systems and aerocapture, and a low-cost aerocapture demonstration [13, 14].

Figure 6 (left) shows the trade space for Jupiter entry using rigid aeroshells. The vehicle design parameters are chosen from the Galileo aeroshell (beta = 267 kg/m², nose radius RN = 0.22 m). Jupiter's enormous gravity well results in very high entry speeds. For example, the Galileo probe entry highlighted in Fig. 6 entered the atmosphere with a relative speed of 47.4 km/s, subjecting it a peak deceleration of over 200g, a peak heat rate of approximately 30,000 W/cm², and a heat load of 200 kJ/cm². The extreme conditions required 50% of the entry mass being TPS, the highest TPS mass fraction for any entry probe. As seen in Fig. 6, Jupiter entry results in heat rates in the range of 30,000 – 100,000 W/cm2 and heat loads in the range of 100s of kJ/cm². The carbon-phenolic TPS material used on the Galileo probe is no longer available due to lack of raw materials, there are no alternative TPS materials that can withstand such extreme conditions precluding another Jupiter entry probe mission in the foreseeable future.

Figure 6 (right) shows the trades space for Saturn entry with a Galileo-like probe. Saturn's large gravity well also results in high entry speeds, though much less than at Jupiter. Comparing the conditions to Jupiter, it is seen that Saturn entry conditions are not as severe as Jupiter and with heat rates in the range of a 1000-3000 W/cm².

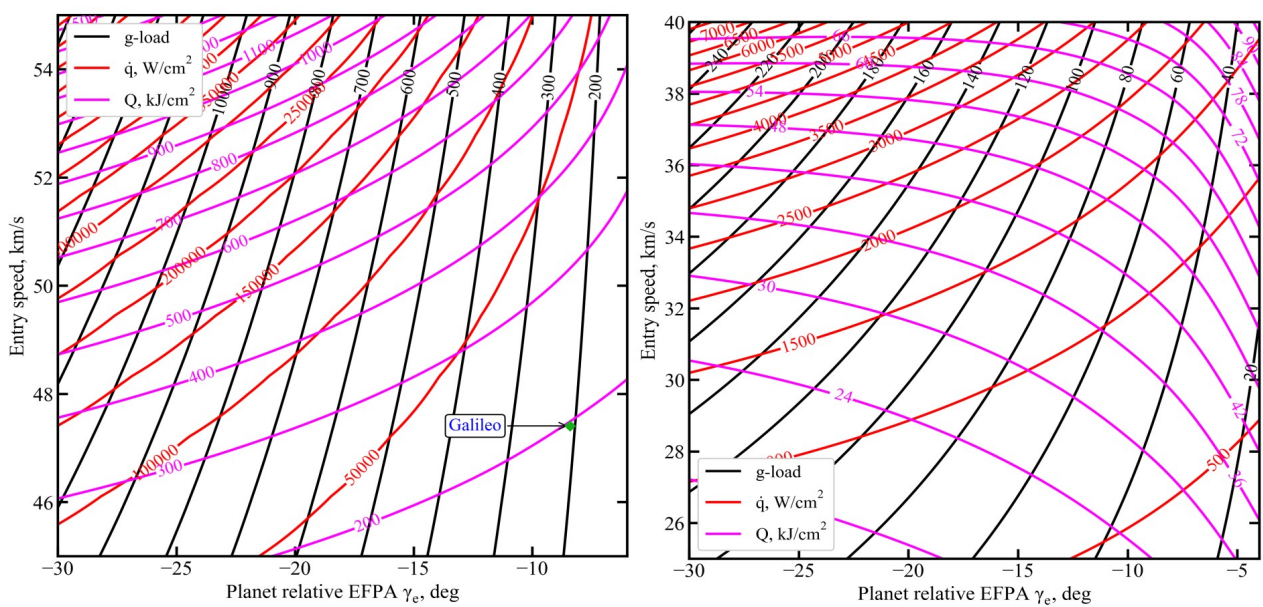

Figure 6. (Left) Trade space for Jupiter entry with rigid aeroshell. (Right) Saturn entry.

Figure 7 shows the trade space for Titan entry. The vehicle design parameters are chosen from the Huygens aeroshell (beta = 35 kg/m², nose radius RN = 1.25 m). Titan's low gravity results in low entry speeds, and the thick extended atmosphere results in low deceleration and heating rates. This has applications for a landers delivered to Titan surface using deployable systems, as well as a future Titan orbiter using aerocapture [15].

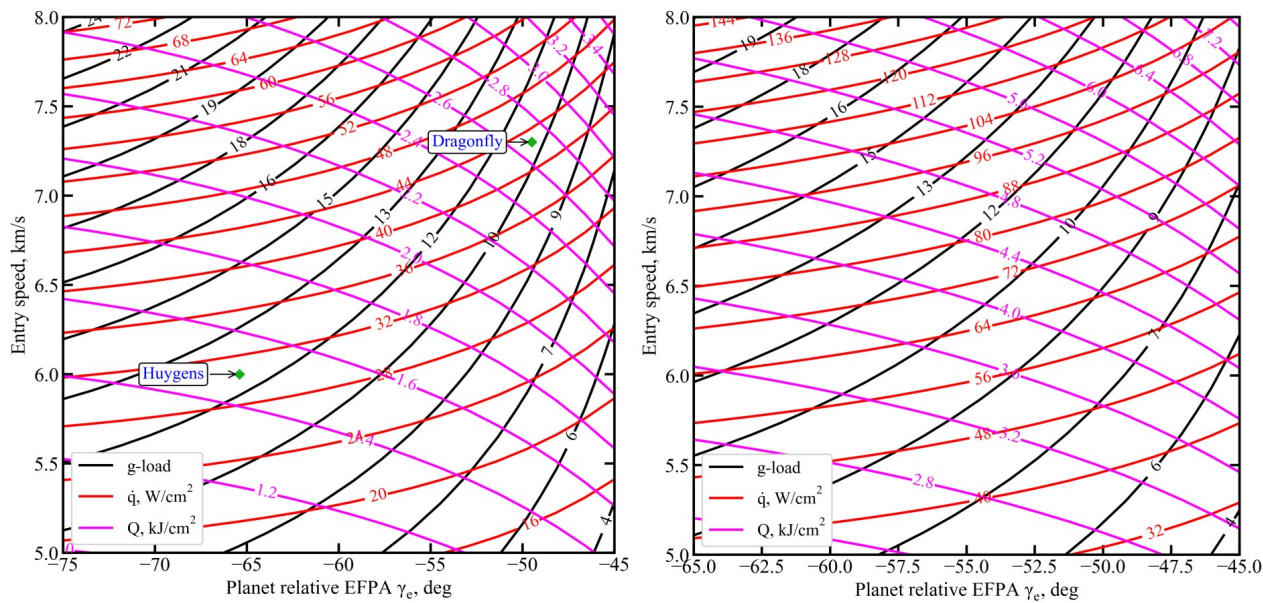

Figure 7. (Left) Trade space for Titan entry with rigid aeroshell, and (right) low b.c.(beta = 25 kg/m²).

Figure 8 shows the entry trade space for Uranus and Neptune using PV heritage aeroshells. The entry speeds are in the range of 20 − 30 km/s, resulting in peak heat rates in the range of 1000 − 5000 W/cm².

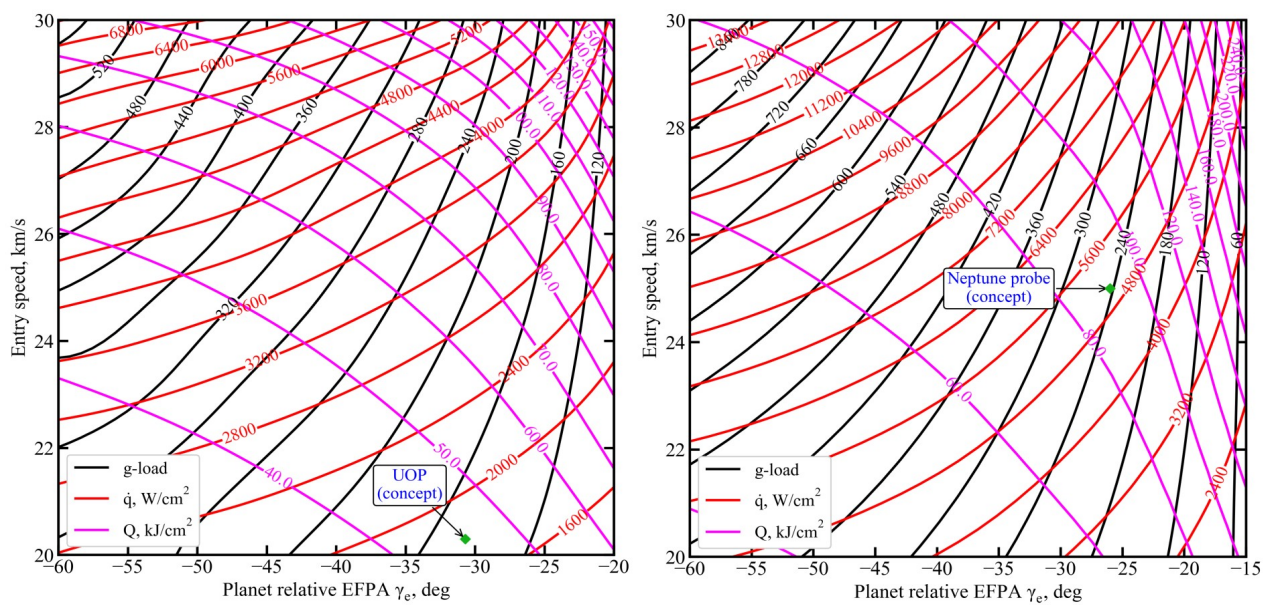

Figure 8. (Left) Trade space for Uranus entry with rigid aeroshell. (Right) Neptune entry.

The proposed Uranus and Probe (UOP) mission for the upcoming decade is highlighted, along with a Neptune probe from the Ice-Giants Mission Study. The high entry speeds result in more demanding aerothermal environments at Uranus and Neptune than at Earth, Mars, or Titan; but is actually comparable to Venus entry. The peak heat rate for aerocapture at Uranus and Neptune is less than for probe entry, as the aerocapture is shallower than that shown in Fig. 8, but the heat loads will be higher compared to probe entry. Nevertheless, the aero-thermal conditions of up to 5000 W/cm$^2$ an be accommodated by existing thermal protection systems such as HEEET. While aerocapture is not considered for the baseline UOP missions [16, 17], aerocapture can enable shorter flight times than possible with propulsive insertion [18, 19], insert more mass into orbit around both Uranus and Neptune for future missions [20, 21].

The analysis presented in the study provides a quick overview of the deceleration and aero-heating environments at all Solar System destinations. Given a flight-path angle and an entry speed, the expected entry conditions can be quickly obtained. The trade space carpet plots presented in the study, along with benchmarks from historical missions, can serve as a useful reference for designing future entry probe and lander mission concepts across the Solar System. Potential applications include small rigid entry probes to Venus [22], Mars sample return mission [23], deployable entry systems for Venus and Mars landers [24], and entry probes at Saturn, Uranus, and Neptune [25].

### III. CONCLUSIONS

This study presented a survey of the design trade space for atmospheric entry missions across the Solar System using carpet plots. Past Venus entry missions have used rigid aeroshells and steep entry to limit the heat load, but subjected them to very high deceleration and peak heat rates. Deployable low-ballistic coefficient systems offer a promising alternative for future Venus entry missions. The entry environment at Venus is less severe than at Earth, and the Mars entry environment is much less demanding than that at Earth. Jupiter presents an extreme environment for entry due to its enormous gravity well which results in very high entry speeds. Saturn entry presents a significantly less demanding environment than Saturn, and is comparable to Venus. Titan's low gravity results in low entry speeds, and the thick extended atmosphere results in low deceleration and heating rates  Uranus and Neptune entry also result in high entry speeds and high aero-thermal loads, but is within the capability of existing thermal protection system materials. The trade space carpet plots presented in the study, along with benchmarks from historical missions, can serve as a useful reference for designing future entry probe and lander mission concepts across the Solar System.

### DATA AVAILABILITY

The trade space charts were produced using the open-source Aerocapture Mission Analysis Tool (AMAT) v2.2.22.  The data presented in the paper will be made available by the author upon reasonable request.

# REFERENCES

[1] National Academies of Sciences, Engineering, and Medicine. 2022. *Origins, Worlds, and Life: A Decadal Strategy for Planetary Science and Astrobiology 2023-2032*. Washington, DC: The National Academies Press. https://doi.org/10.17226/26522

[2] NASA Ames Research Center, Entry Systems and Technology Division. Planetary Mission Entry Vehicle Quick Reference Guide. NASA, 2022. NASA/SP-20220010761.

[3] Girija AP, "Comparative Study of Planetary Atmospheres and Implications for Atmospheric Entry Missions," arXiv, Vol. 2307, No. 16277, 2023, pp. 1-15
https://doi.org/10.48550/arXiv.2307.16277

[4] Girija AP et al., "A Unified Framework for Aerocapture Systems Analysis," *AAS/AIAA Astrodynamics Specialist Conference,* 2019, pp 1-21.
https://doi.org/10.31224/osf.io/xtacw

[5] Girija AP et al. "Quantitative assessment of aerocapture and applications to future solar system exploration." *Journal of Spacecraft and Rockets,* Vol. 59, No. 4, 2022, pp. 1074-1095.
https://doi.org/10.2514/1.A35214

[6] Girija AP, "A Systems Framework and Analysis Tool for Rapid Conceptual Design of Aerocapture Missions," Ph.D. Dissertation, Purdue University Graduate School, 2021.
https://doi.org/10.25394/PGS.14903349.v1

[7] Girija AP et al. "AMAT: A Python package for rapid conceptual design of aerocapture and atmospheric Entry, Descent, and Landing (EDL) missions in a Jupyter environment," *Journal of Open Source Software,* Vol. 6, No. 67, 2021, pp. 3710.
https://doi.org/10.21105/joss.03710

[8] Girija AP, Lu Y, and Saikia SJ, "Feasibility and mass-benefit analysis of aerocapture for missions to Venus," *Journal of Spacecraft and Rockets,* Vol. 57, No. 1, 2020, pp. 58-73.
https://doi.org/10.2514/1.A34529

[9] Austin A et al., "Enabling and Enhancing Science Exploration Across the Solar System: Aerocapture Technology for SmallSat to Flagship Missions," *Bulletin of the American Astronomical Society,* Vol. 53, No. 4, 2021, pp. 057.
https://doi.org/10.3847/25c2cfeb.4b23741d

[10] Girija AP, Saikia SJ, and Longuski JM, "Aerocapture: Enabling Small Spacecraft Direct Access to Low-Circular Orbits for Planetary Constellations," *Aerospace,* Vol. 10, No. 3, 2023, pp. 271.
https://doi.org/10.3390/aerospace10030271

[11] Limaye SS et al., "Venus observing system," *Bulletin of the American Astronomical Society,* Vol. 53, No. 4, 2021, pp. 370.
https://doi.org/10.3847/25c2cfeb.7e1b0bf9

[12] Shibata E et al., "A Venus Atmosphere Sample Return Mission Concept: Feasibility and Technology Requirements," *Planetary Science Vision 2050 Workshop,* 2017, pp. 8164.

[13] Girija AP, "Aerocapture: A Historical Review and Bibliometric Data Analysis from 1980 to 2023," arXiv, Vol. 2307, No. 01437, 2023, pp 1-19.
https://doi.org/10.48550/arXiv.2307.01437

[14] Girija AP, "A Low Cost Mars Aerocapture Technology Demonstrator," arXiv, Vol. 2307, No. 11378, 2023, pp. 1-14
https://doi.org/10.48550/arXiv.2307.11378

[15] Girija AP, "ADEPT Drag Modulation Aerocapture: Applications for Future Titan Exploration," arXiv, Vol. 2306, No. 10412, 2023, pp 1-27.
https://doi.org/10.48550/arXiv.2306.10412

[16] Cohen I et al., "New Frontiers-class Uranus Orbiter: Exploring the feasibility of achieving multidisciplinary science with a mid-scale mission," Bulletin of the American Astronomical Society, Vol. 53, No. 4, 2021, pp. 323.
https://doi.org/10.3847/25c2cfeb.262fe20d

[17] Jarmak S et al., "QUEST: A New Frontiers Uranus orbiter mission concept study," Acta Astronautica, Vol. 170, 2020, pp. 6-26.
https://doi.org/10.1016/j.actaastro.2020.01.030

[18] Girija AP, et al., "Feasibility and performance analysis of neptune aerocapture using heritage blunt-body aeroshells," Journal of Spacecraft and Rockets, Vol. 57, No. 6, 2020, pp. 1186-1203.
https://doi.org/10.2514/1.A34719

[19] Girija AP, "A Flagship-class Uranus Orbiter and Probe mission concept using aerocapture," Acta Astronautica Vol. 202, 2023, pp. 104-118.
https://doi.org/10.1016/j.actaastro.2022.10.005

[20] Dutta S et al., "Aerocapture as an Enhancing Option for Ice Giants Missions," Bulletin of the American Astronomical Society, Vol. 53, No. 4, 2021, pp. 046.
https://doi.org/10.3847/25c2cfeb.e8e49d0e

[21] Iorio L et al., "One EURO for Uranus: the Elliptical Uranian Relativity Orbiter mission." Monthly Notices of the Royal Astronomical Society, Vol. 523, No. 3, 2023, pp. 3595-3614
https://doi.org/10.1093/mnras/stad1446

[22] French R et al., "Rocket lab mission to venus," Aerospace, Vol. 9, No. 8, 2022, pp. 445.
https://doi.org/10.3390/aerospace9080445

[23] Muirhead BK et al. "Mars sample return mission concept status," 2020 IEEE Aerospace Conference, IEEE, 2020.
https://doi.org/10.1109/AERO47225.2020.9172609

[24] Woolley R et al. "Rideshare Strategies for Small Mars Missions." 2021 IEEE Aerospace Conference, IEEE, 2021, pp. 1-8.
https://doi.org/10.1109/AERO50100.2021.9438174

[25] Beddingfield C et al., "Exploration of the Ice Giant Systems," Bulletin of the American Astronomical Society, Vol. 53, No. 4, 2021, pp. 121.
https://doi.org/10.3847/25c2cfeb.e2bee91e